\documentclass[a4paper,11pt]{article}
\usepackage{pos}

\newcommand{\mi}{\mathrm{i}} 

\newcommand{\dx}[1]{\hspace{-0.4em}\ensuremath{\mathrm{d}#1}\,}

\newcommand{\eqn}[1]{Eq.~(\ref{#1})}
\newcommand{\fig}[1]{Fig.~\ref{#1}}
\newcommand{\tab}[1]{Table~\ref{#1}}

\title{\vspace{-25mm}
  \begin{flushright}
    LFTC-21-7/68. 
  \end{flushright}
  \vspace{15mm}
  Hidden charm mesons  in nuclear matter and nuclei}
  \ShortTitle{Hidden charm mesons  in nuclear matter and nuclei}

\author*[a]{J.J.~Cobos Mart\'inez}
\author[b]{K.~Tsushima}
\author[c]{G.~Krein}
\author[d]{A.W.~Thomas}

\affiliation[a]{Departamento de F\'isica, Universidad de Sonora, Boulevard
Luis Encinas J. y Rosales, Colonia Centro, Hermosillo, Sonora 83000, M\'exico}
\affiliation[b]{Laborat\'orio de F\'{\i}sica Te\'orica e Computacional-LFTC, 
Universidade Cidade de S\~ao Paulo,
01506-000, S\~ao Paulo, SP, Brazil}
\affiliation[c]{Instituto de F\'{\i}sica Te\'orica, Universidade Estadual 
Paulista, Rua Dr. Bento Teobaldo Ferraz, 271 - Bloco II, 01140-070, 
S\~ao Paulo, SP, Brazil}
\affiliation[d]{CSSM, School of Physical Sciences, University of Adelaide,
Adelaide SA 5005, Australia}

\emailAdd{jesus.cobos@fisica.uson.mx}
\emailAdd{kazuo.tsushima@gmail.com}
\emailAdd{gkrein@ift.unesp.br}
\emailAdd{anthony.thomas@adelaide.edu.au}

\abstract{Recent results for the $\eta_c$- and $J/\psi$-nucleus bound state energies for various nuclei are presented. 
The attractive potentials for the $\eta_c$ and $J/\psi$ mesons in the nuclear medium originate, respectively, from the in-medium enhanced $DD^{*}$ and $D\bar{D}$ loops in the $\eta_c$ and $J/\psi$ self energies. Our results suggest that the $\eta_c$ and $J/\psi$ mesons should form bound states with all the nuclei considered}

\FullConference{%
  *** 22nd Particles and Nuclei International Conference (PANIC 2021), ***\\
  *** 5-10 September, 2021 ***\\
  *** Lisbon, Portugal - Online ***
}

\begin{document}
\maketitle

\section{Introduction}

The study of the interactions between charmonium states, such as $\eta_c$ and $J/\Psi$,  and 
atomic nuclei offers an important opportunity to extend our knowledge of the strong force and  strongly  
interacting matter.
Since charmonia and nucleons do not share light quarks, the Zweig rule suppresses interactions
mediated by the exchange of mesons made of light quarks. Thus, it is important to explore other
mechanisms that could lead to the binding  between charmonia and atomic nuclei.

Recent studies~\cite{JPsiMassAndBS} have shown that modifications induced by the strong nuclear mean fields on the $D$ mesons' light-quark content enhance the $J/\Psi$
self-energy in such that there is an attractive $J/\Psi$-nucleus effective potential. 
Here we update our previous study on the  $J/\Psi$~\cite{JPsiMassAndBS} and extend it to the case of 
$\eta_c$~\cite{Cobos-Martinez:2020ynh} by considering an intermediate $DD^{*}$ state in the $\eta_c$ self-energy with the light quarks created  from the vacuum.
Furthermore, recently, we have extended this approach to the case of bottomonium~\cite{Zeminiani:2020aho}.

\section{\label{sec:mass_shifts}$\eta_c$ and $J/\Psi$ scalar potentials in symmetric nuclear matter}

For the computation of the $\eta_c$ and $J/\Psi$ scalar potentials in nuclear matter we use an
effective Lagrangians approach at the hadronic level--see Ref.~\cite{Cobos-Martinez:2020ynh} and referenes therein. The interaction Lagrangians for the $\eta_c D D^{*}$ and $J/\Psi D D$ vertices are
given by
\begin{equation}
\label{eqn:Leff}
\mathcal{L}_{\eta_c D D^{*}}=\mi g_{\eta_c}(\partial_{\mu} \eta_c)
\left[\bar{D}^{*\mu}\cdot D-\bar{D}\cdot D^{*\mu}\right], \hspace{3mm}
\mathcal{L}_{\psi D D}=\mi g_{\psi}\psi^{\mu}
\left[\bar{D}\cdot (\partial_{\mu}D)-(\partial_{\mu}\bar{D})
\cdot D\right] \, ,
\end{equation}
where we have denoted the $J/\Psi$ vector field by $\psi$, $D^{(*)}$ represents an isospin doublet, and $g_{\eta_c}$ and $g_{\psi}$ are, respectively the  coupling constants for
the vertices $\eta_c D D^{*}$ and $J/\Psi D \bar{D}$.

We use an effective Lagrangian, such as  \eqn{eqn:Leff}, to compute the $\eta_c$ and 
$J/\Psi$ scalar potentials in nuclear matter and nuclei,  following our previous  
works~\cite{JPsiMassAndBS,PhiMassAndBS}.  
In explaining our approach, we focus on the $\eta_c$ meson~\cite{Cobos-Martinez:2020ynh} but 
similar expressions can be obtained for the $J/\Psi$--see Refs.~\cite{JPsiMassAndBS};
however, we will show the correspondig results for the $J/\Psi$. 		

The starting point is the computation of the $\eta_c$ mass shift in nuclear matter, which is given
by $\Delta m_{\eta_c}= m_{\eta_c}^{*}-m_{\eta_c}$, where $m_{\eta_c}^{*}$ ($m_{\eta_c}$) is the 
in-medium (vacuum) mass of the $\eta_c$. 
These masses are obtained by solving 
$m_{\eta_c}^{2}= (m_{\eta_c}^{0})^{2} + \Sigma_{\eta_c}(k^{2}=m_{\eta_c}^{2})$,
where $m_{\eta_c}^{0}$ is the bare $\eta_c$ mass and $\Sigma_{\eta_c}(k^{2})$ is 
the $\eta_c$ self-energy, which is calculated for a $\eta_c$ at rest and considering only the $DD^{*}$ intermediate state, and is given by 
$\Sigma_{\eta_c}(k^{2})= (8g_{\eta_c D D^{*}}^{2}/\pi^{2})\int_{0}^{\infty}\dx{k}k^{2}I(k^{2})$.

The integreal $I(k^2)$, given in Eq.~(3) of Ref.~\cite{Cobos-Martinez:2020ynh}, is divergent and will be
 regulated with a phenomenological vertex form factor,  as in 
 Refs~\cite{JPsiMassAndBS, Cobos-Martinez:2020ynh, Zeminiani:2020aho,PhiMassAndBS}, with 
 cutoff parameter $\Lambda_{D}\,(=\Lambda_{D^{*}})$, which we vary in the  range 1500-3000 MeV, 
 for both the $\eta_c$ and $J/\Psi$, to quantify the sensitivity of our results to its value.
 For the coupling constants we use the values $g_{\eta_c}=(0.6/\sqrt{2})g_{J/\Psi}=3.24$ 
 and $g_{J/\Psi}=7.64$~\cite{Lin:1999ad,Lucha:2015dda}.
 The $\eta_c$ mass in the nuclear medium is computed  with the  self-energy calculated with the medium-modified $D$ and $D^{*}$ masses~\cite{Cobos-Martinez:2020ynh}.
\begin{figure*}
\scalebox{0.64}{
\begingroup
\setlength{\tabcolsep}{1pt} 
\renewcommand{\arraystretch}{0.1} 
\begin{tabular}{lll}
\includegraphics[scale=0.3]{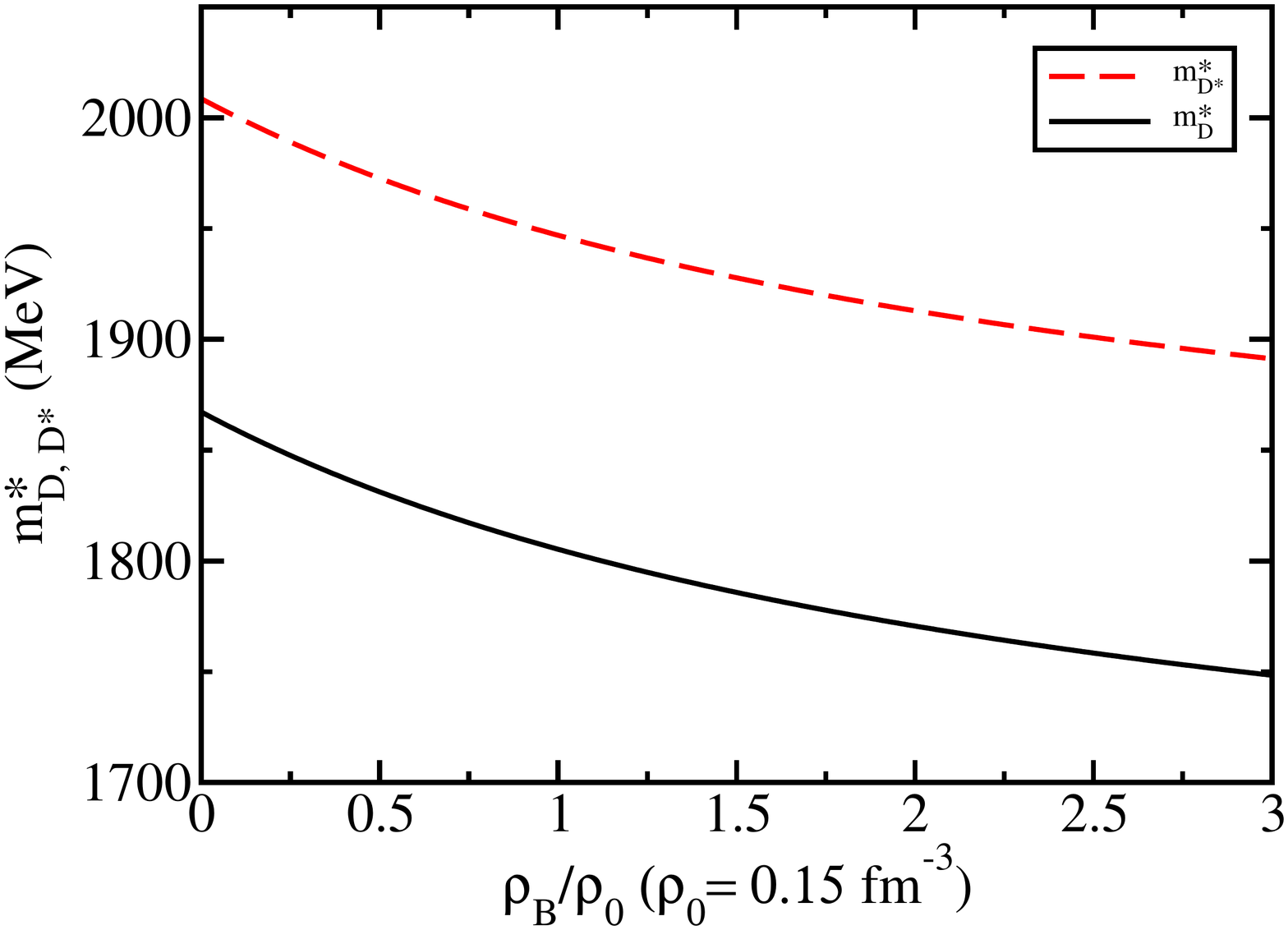} &
\includegraphics[scale=0.3]{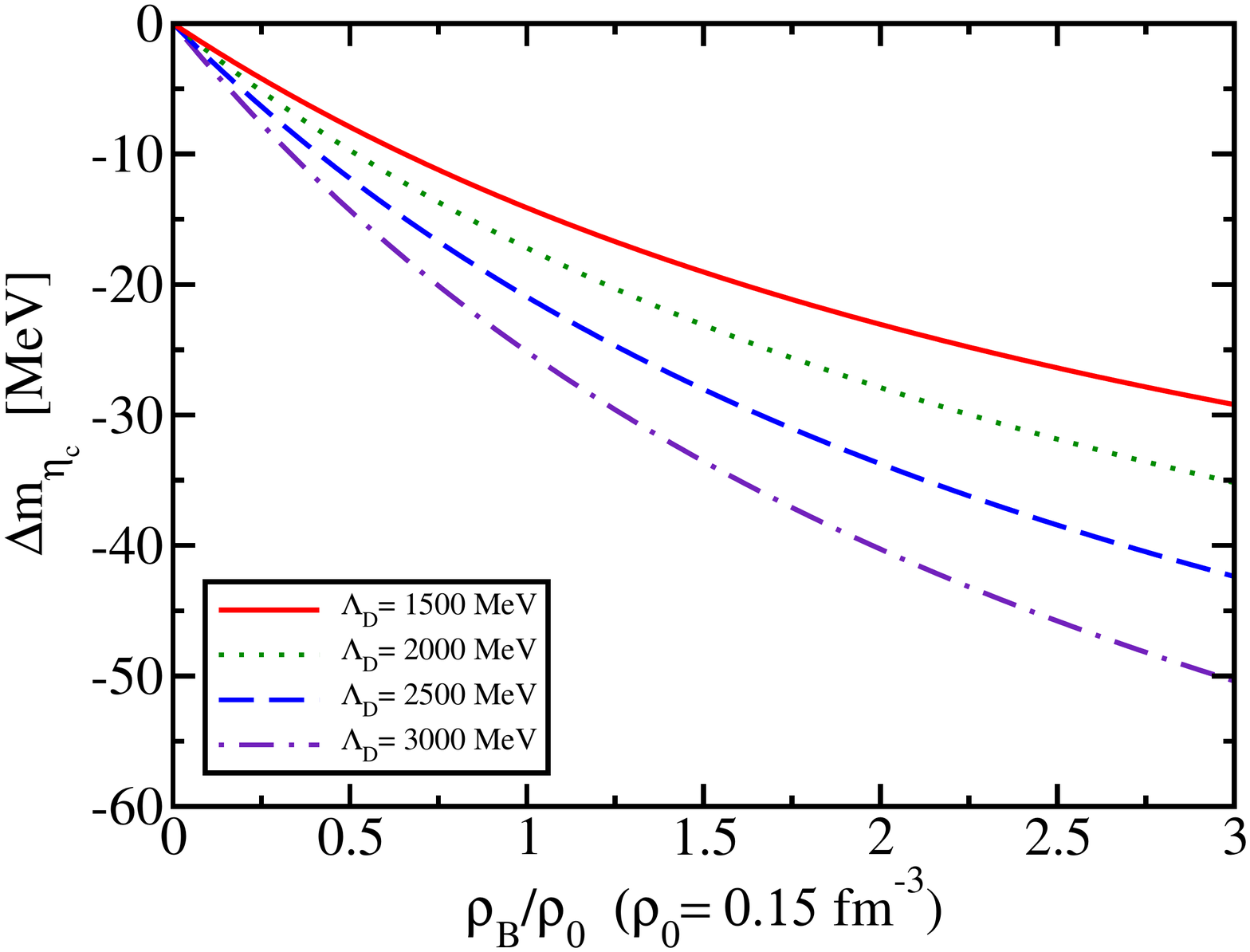} &
\includegraphics[scale=0.3]{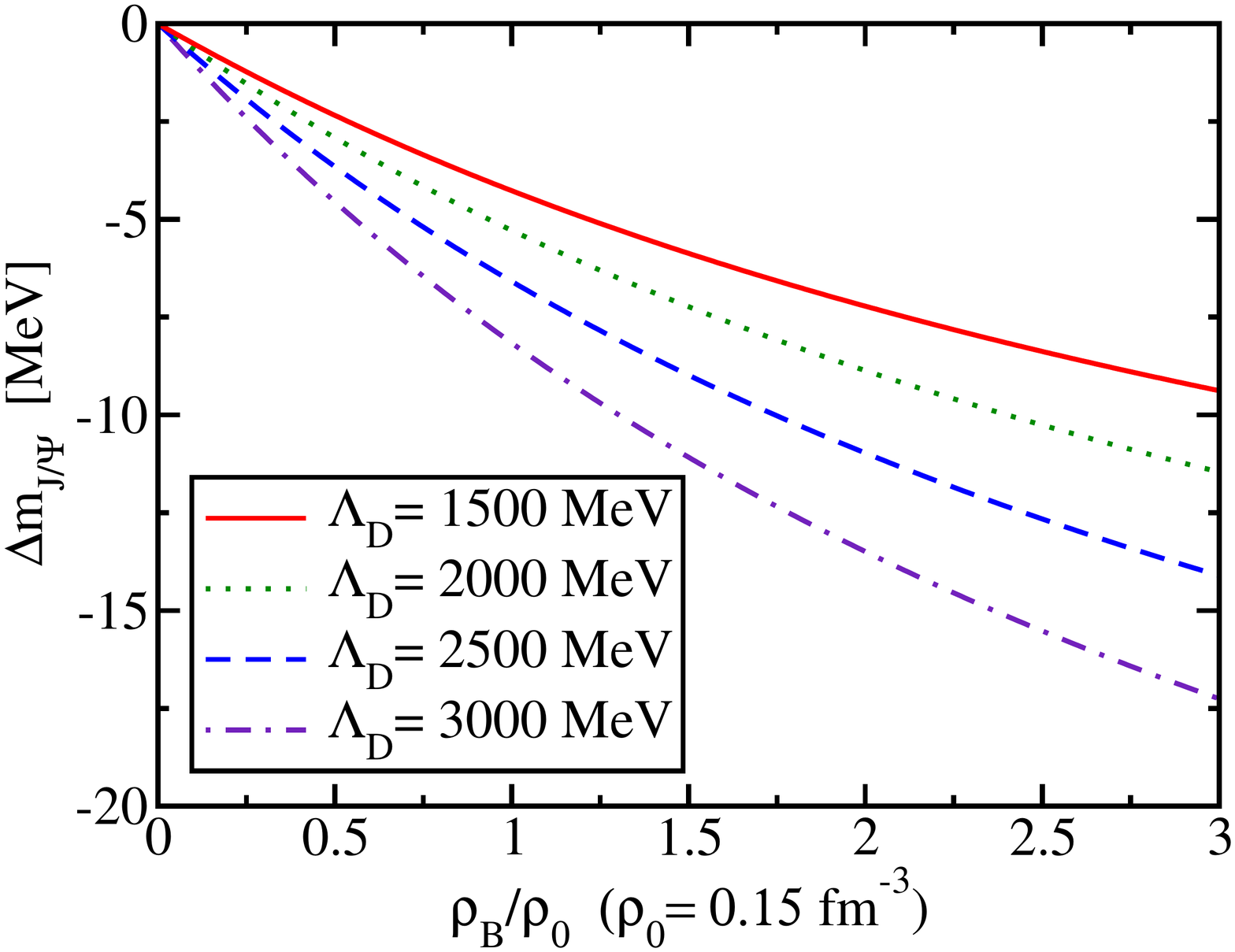} \\
\hspace{43mm}(a) & \hspace{43mm}(b) & \hspace{43mm}(c) 
  \end{tabular}
  \endgroup
  }
  \caption{\label{fig:masses}
    $D$ and $D^{*}$ masses and  $\eta_c$ and $J/\Psi$ mass shifts in nuclear matter.}
\end{figure*}
%
The nuclear density dependence of the $\eta_c$ ($J/\Psi$) mass originates from the interactions of the
intermediate  $DD^{*}\,(D\bar{D})$ state with the nuclear medium through their medium-modified
 masses, $m_{D}^{*}$ and $m_{D^{*}}^{*}$. These masses, calculated within
 the quark-meson  coupling (QMC) model~\cite{JPsiMassAndBS}, are shown in 
 \fig{fig:masses} as a function of $\rho_{B}/\rho_{0}$, where $\rho_{B}$ is the baryon 
density of nuclear matter and  $\rho_{0}=0.15$ fm$^{-3}$ is the  saturation density of symmetric 
nuclear  matter. As can be seen from \fig{fig:masses} the masses of the $D$ and $D^{*}$ mesons
decrease in the nuclear medium. 

In \fig{fig:masses} we present the $\eta_c$ mass shift in nuclear matter. 
Also, in \fig{fig:masses} we show our updated corrresponding results for the $J/\Psi$ mass
shift.
As can be seen from \fig{fig:masses} the effect of the nuclear medium on the  $\eta_c$ and $J/\Psi$
is to provide attraction to these mesonsm and this  could potentially lead to the binding of these 
mesons to nuclei since, in each case, the mass shift is significant for $\rho_{B}/\rho_{0}\sim 1$.

\section{\label{sec:etac_bound_states} Charmonium-nucleus bound states}


\begin{figure}
\scalebox{0.64}{
\begingroup
\setlength{\tabcolsep}{1pt} 
\renewcommand{\arraystretch}{0.1} 
\begin{tabular}{lll} 
\includegraphics[scale=0.3]{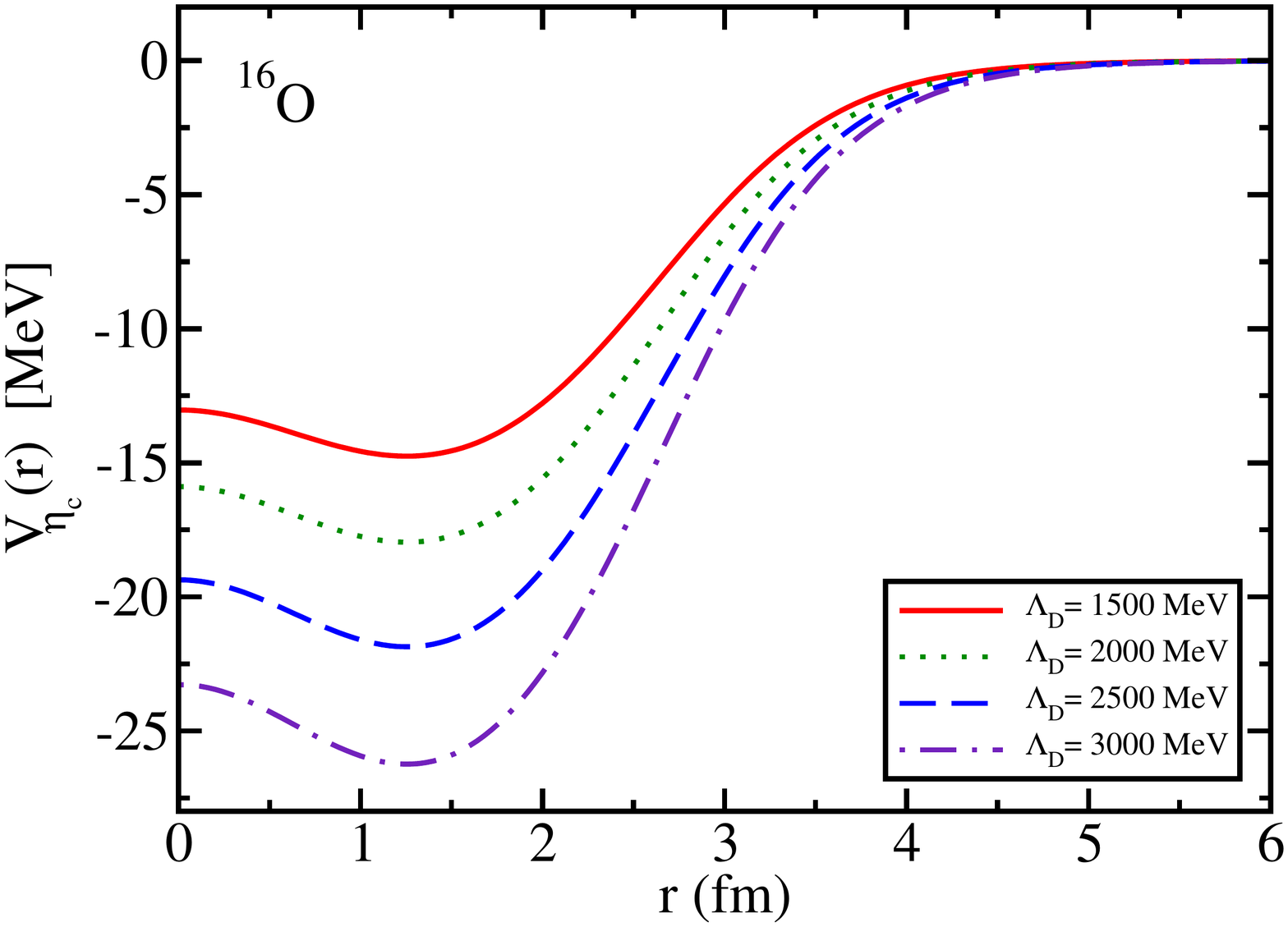} &
\includegraphics[scale=0.3]{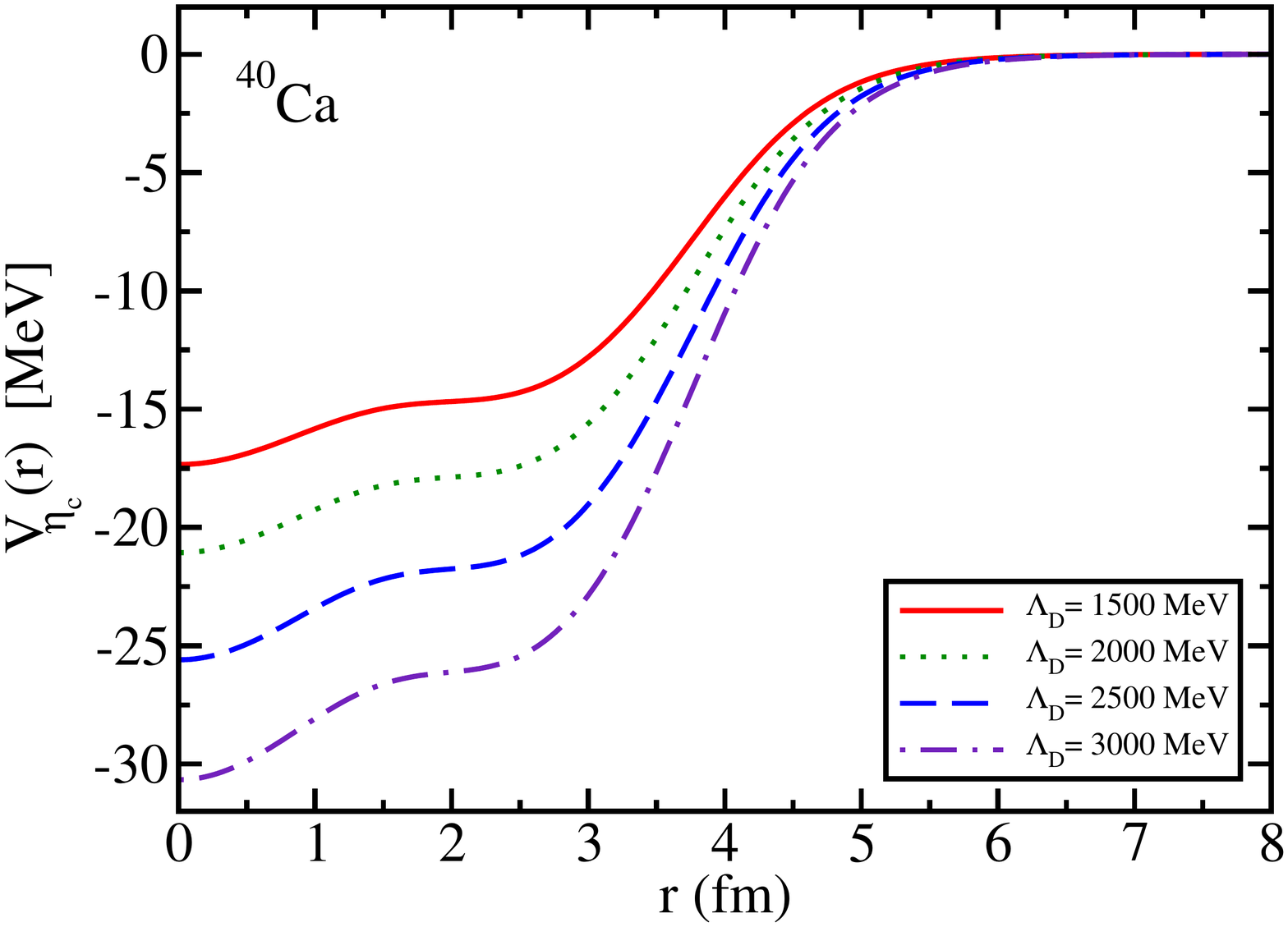}  &
\includegraphics[scale=0.3]{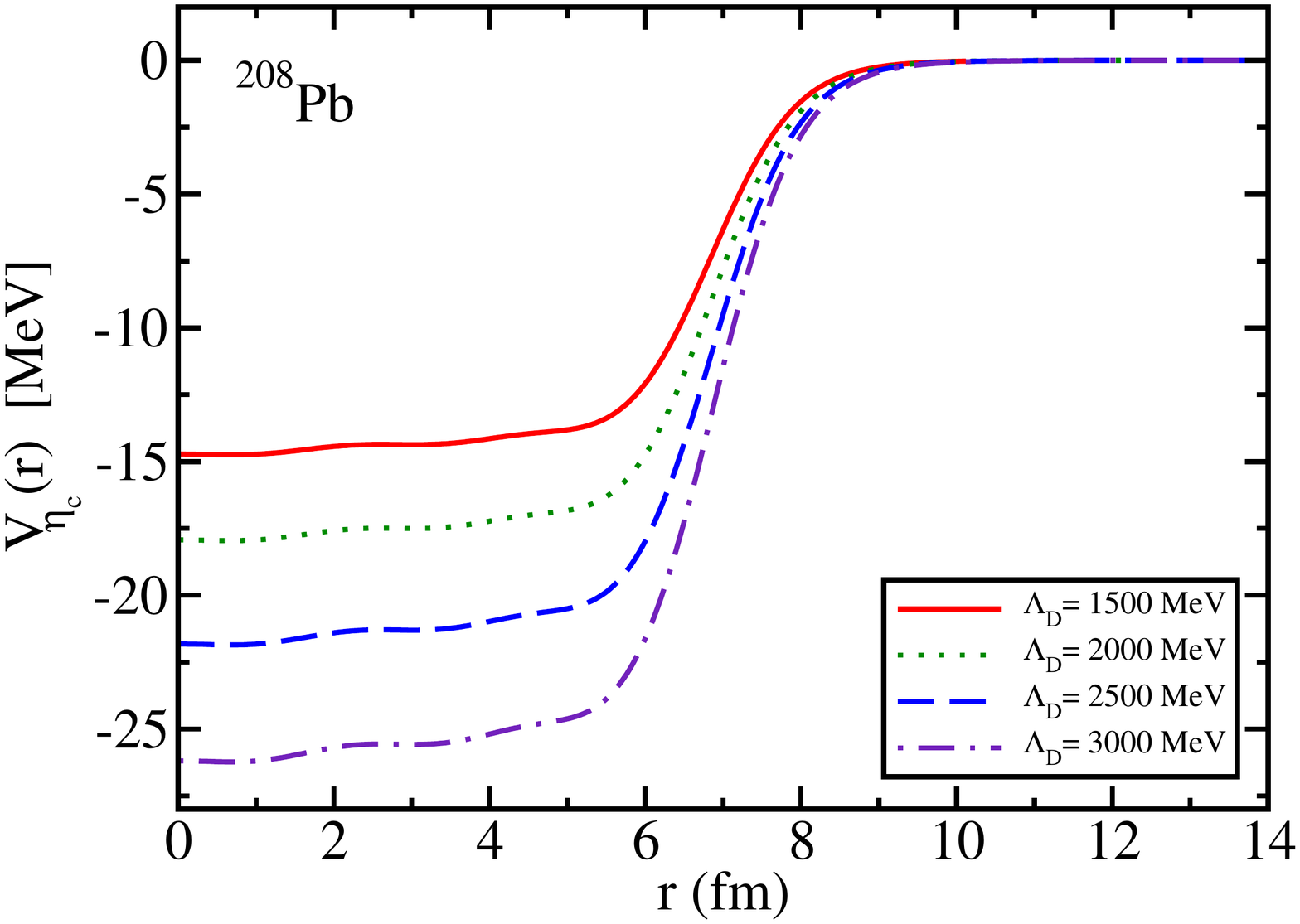}  \\
\includegraphics[scale=0.3]{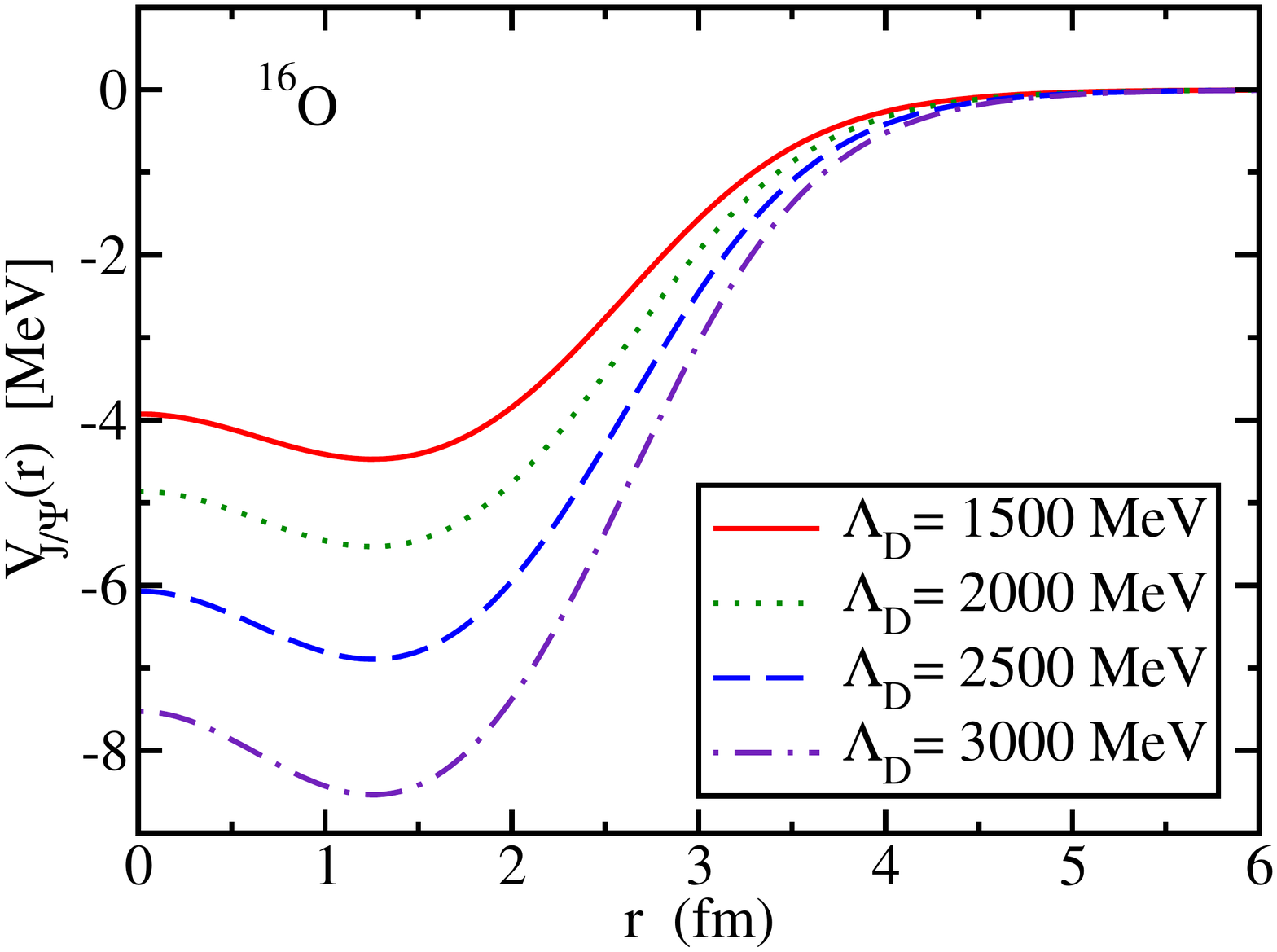} &
\includegraphics[scale=0.3]{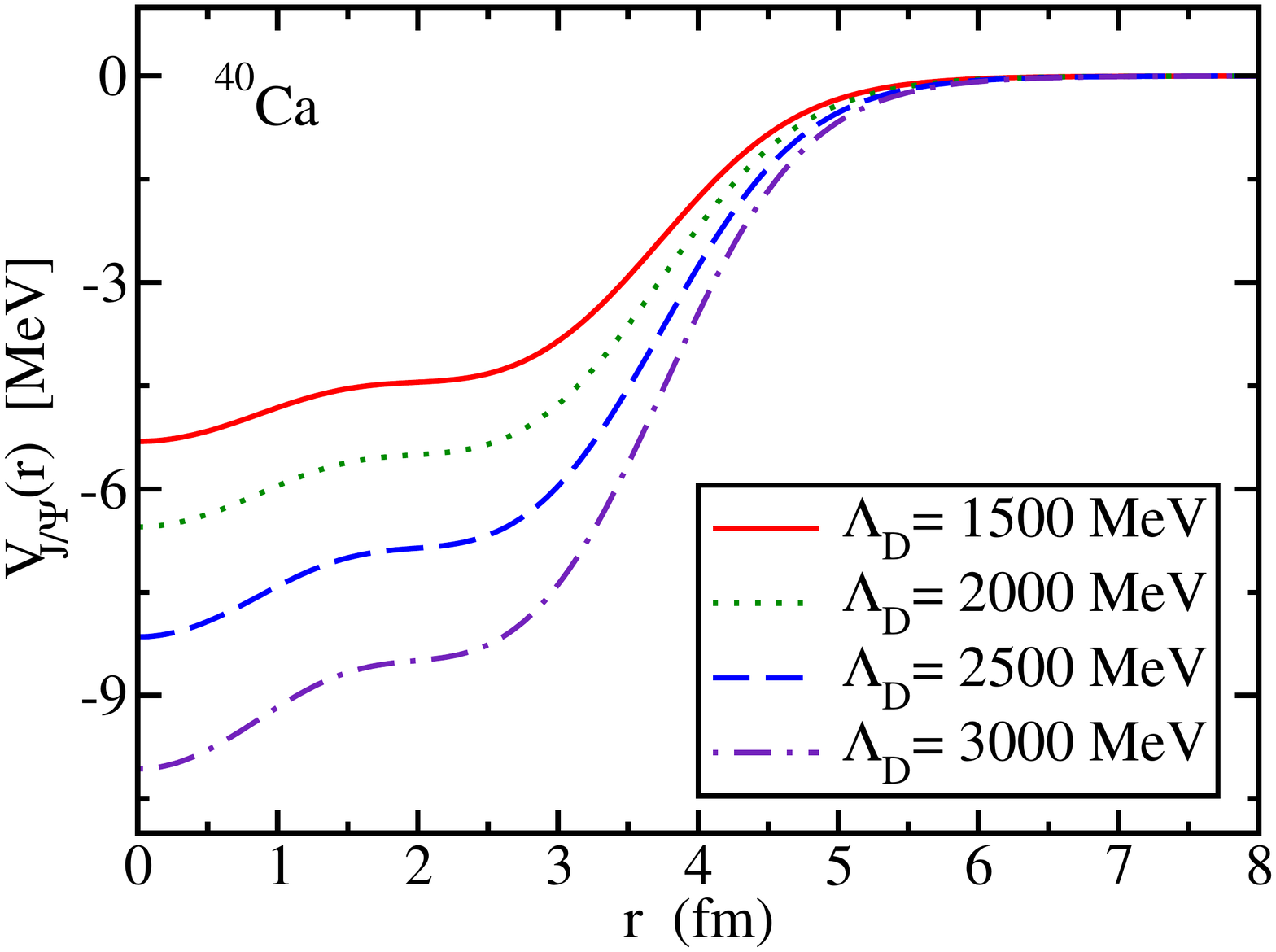} &
\includegraphics[scale=0.3]{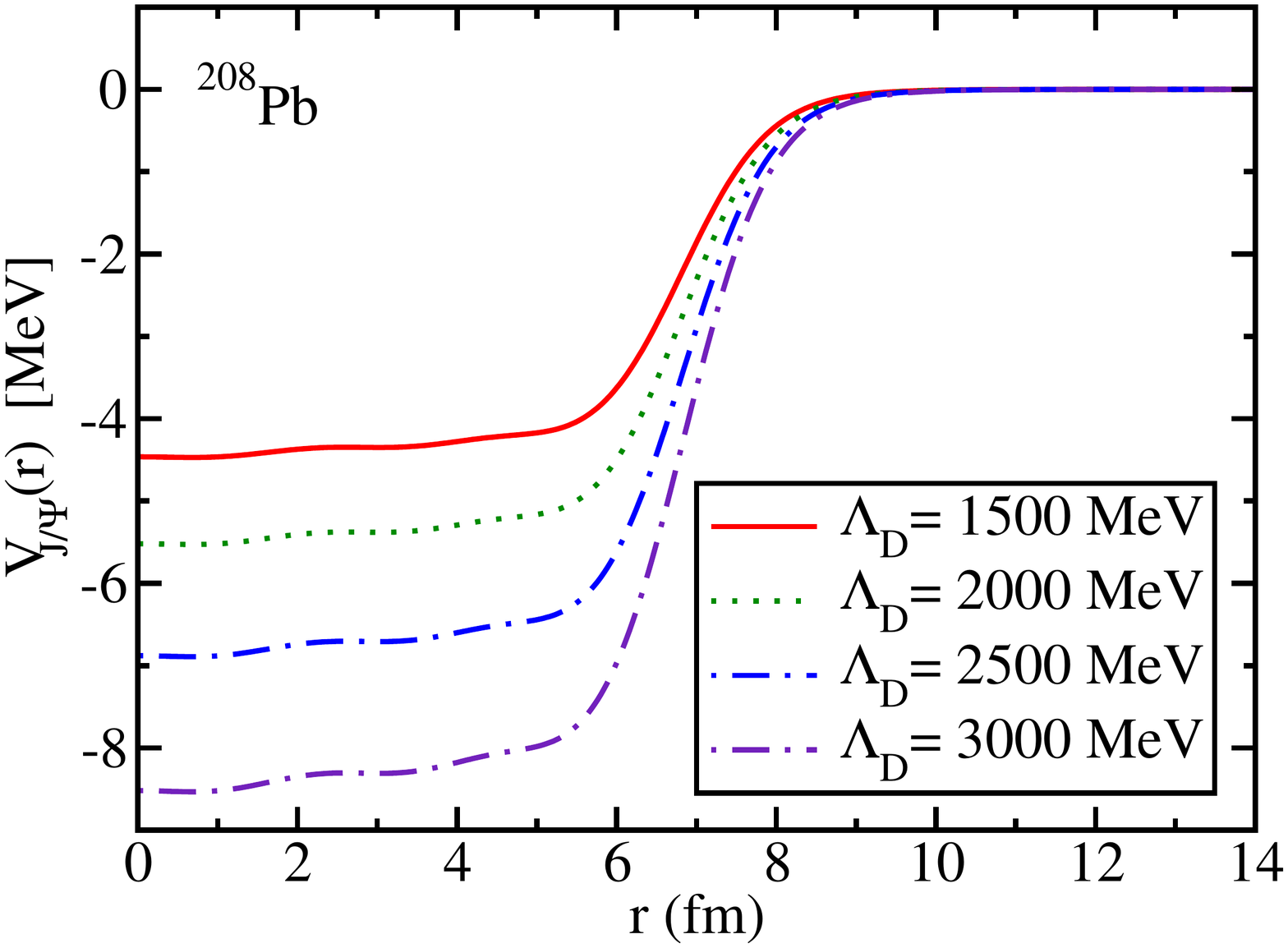}
  \end{tabular}
  \endgroup
  }
  \caption{\label{fig:potentials}
    $\eta_c$- and $J/\Psi$-nucleus  potentials for various nuclei and values of the cutoff parameter $\Lambda_{D}$.}
\end{figure}

We now consider the binding of the $\eta_c$ and $J/\Psi$ mesons to nuclei when these mesons are produced inside nucleus $A$ with baryon density distribution $\rho_{B}^{A}(r)$; see  Ref.~\cite{Cobos-Martinez:2020ynh}.
The nuclear density distribution for the nuclei we consider here are calculated within the 
QMC model, except for $^{4}$He, whose parametrization was obtained in Ref.~\cite{Saito:1997ae}.

\subsection{\label{sec:etac} $\eta_c$-nucleus bound states}

Using a local density approximation, the $\eta_c$-meson potential  inside nucleus $A$ is given by 
$V_{\eta_c A}(r)= \Delta m_{\eta_c}(\rho_{B}^{A}(r))$, where $\Delta m_{\eta_c}$ is the $\eta_c$ mass 
shift shown in \fig{fig:masses}, and  $r$ is the distance from the center of the nucleus. 

In~\fig{fig:potentials} we present the $\eta_c$ potentials for the various nuclei.
From \fig{fig:potentials}, we can see that all $\eta_c$ potentials are attractive but their depths 
depend on $\Lambda_{D}$.
Next, we calculate the $\eta_c$--nucleus bound state energies by solving the Klein-Gordon equation
$\left(-\nabla^{2} + m^{2} + 2 m V(\vec{r})\right)\phi_{\eta_c}(\vec{r}) 
= \mathcal{E}^{2}\phi_{\eta_c}(\vec{r})$, where $m$ is the reduced mass of the $\eta_c$--nucleus 
system in vacuum  and $V(\vec{r})$ is the $\eta_c$-nucleus potential shown in \fig{fig:potentials}.

The bound state energies ($E$) of the $\eta_c$-nucleus system, given by $E= \mathcal{E}-m$,  
are listed in  \tab{tab:Energies}.
These results show that the $\eta_c$ is expected to form bound states with all the nuclei
 considered, independently of the value of $\Lambda_D$. Expectedly, the particular values for the 
 bound state energies are dependent on $\Lambda_{D}$.
 This dependence is an uncertainty in the results obtained in our approach.
Finally, note that the $\eta_c$ is predicted to be bound more strongly to heavier nuclei.

\begin{table}
\begin{minipage}{0.5\textwidth}
\scalebox{0.65}{
\begin{tabular}{ll|r|r|r|r}
  \hline \hline
  & & \multicolumn{4}{c}{Bound state energies ($E$)} \\
  \hline
& $n\ell$ & $\Lambda_{D}=1500$ & $\Lambda_{D}=2000$ & $\Lambda_{D}=2500$ & 
$\Lambda_{D}= 3000$ \\
\hline
$^{4}_{\eta_{c}}\text{He}$
    & 1s & -1.49 & -3.11 & -5.49 & -8.55 \\
\hline
$^{16}_{\eta_{c}}\text{O}$
    & 1s & -7.35 & -9.92 & -13.15 & -16.87 \\
    & 1p & -1.94 & -3.87 & -6.48  & -9.63 \\
    \hline
$^{40}_{\eta_{c}}\text{Ca}$
    & 1s & -11.26 & -14.42 & -18.31 & -22.73 \\
    & 1p & -7.19  & -10.02 & -13.59 & -17.70 \\
\hline
$^{208}_{\eta_{c}}\text{Pb}$
    & 1s & -12.99 & -16.09 & -19.82 & -24.12 \\
    & 1p & -11.60 & -14.64 & -18.37 & -22.59 \\
    & 1d & -9.86  & -12.83 & -16.49 & -20.63 \\
\hline 
\hline
\end{tabular}
}
\end{minipage}
\begin{minipage}{0.5\textwidth}
\scalebox{0.65}{
\begin{tabular}{ll|r|r|r|r}
  \hline \hline
  & & \multicolumn{4}{c}{Bound state energies ($E$)} \\
  \hline
& $n\ell$ & $\Lambda_{D}=1500$ & $\Lambda_{D}=2000$ &$\Lambda_{D}=2500$ & 
$\Lambda_{D}=3000$ \\
\hline
$^{4}_{J/\Psi}\text{He}$
& 1s & n & n & -0.01 & -0.02 \\ 
\hline
$^{16}_{J/\Psi}\text{O}$
& 1s & -0.52 & -1.03 & -1.81 & -2.87 \\
\hline
$^{40}_{J/\Psi}\text{Ca}$
& 1s & -1.92 & -2.78 & -3.96 & -5.45 \\
& 1p &     n & -0.38 & -1.18 & -2.32  \\
\hline
$^{208}_{J/\Psi}\text{Pb}$
& 1s & -3.28 & -4.26 & -5.53 & -7.08 \\
& 1p & -2.24 & -3.16 & -4.38 & -5.87 \\
& 1d & -1.02 & -1.84 & -2.97 & -4.38 \\
\hline \hline
\end{tabular}
}
\end{minipage}
  \caption{\label{tab:Energies} $\eta_c$- and $J/\Psi$-nucleus bound state energies for various nuclei. Dimensionful quantities are in MeV.}
\end{table}

\subsection{\label{sec:jpsi} $J/\Psi$-nucleus bound states}

Using a similar approach to the one just described for the $\eta_c$, the $J/\Psi$ mass shift in
nuclear matter and $J/\Psi$--nucleus bound state energies were calculated in 
Ref.~\cite{JPsiMassAndBS}. 
In these studies, the $J/\Psi$ self-energy involved the $D$, $\bar{D}$, $D^{*}$, and $\bar{D}^{*}$
mesons as intermediate states.  However, it turned out that the loops involving the $D^{*}$ and 
$\bar{D}^{*}$ mesons gave a larger contribution to the $J/\Psi$ self-energy, which is unexpected.
The reason for this could be that in those studies the coupling constants for all the vertices were
 assumed to equal to the $J/\Psi D\bar{D}$ coupling and, furthermore,  that the same cutoff 
 ($\Lambda_{D}$) was used in all vertex form factors.  This latter issue, in particular, should play 
 an important role, as heavier intermediate states induce fluctuations from shorter distances and 
 a corresponding readjustment of the  cutoffs  might be required to compensate for such fluctuations.  
 These issues call for further investigations in the future. The following updated results 
are obtained by considering only the lightest intermediate state in the $J/\Psi$ self-energy, namely 
the $D\bar{D}$ loop.

In \fig{fig:masses} we present results for the $J/\Psi$ mass shift in nuclear matter.
These results show an attractive potential scalar potential for the $J/\Psi$ in nuclear matter in all cases. 
In \fig{fig:potentials} we present the $J/\Psi$ potentials in nuclei, from which we can see,
 as in the $\eta_c$ case, that these are all attractive but their depths are sensitive to the value  
 of the cutoff parameter. Note that the $\eta_c$ potentials are more attractive than those of the
  $J/\Psi$.
The bound state energies for the $J/\Psi$ for various nuclei are listed in \tab{tab:Energies}.
These results show that the $J/\Psi$ is expected to form bound states for all the cases
considered, but only in some cases for $^{4}$He.
Thus, it will be possible to search  for the bound states in a $^{208}$Pb nucleus at JLab now that 
the 12 GeV upgrade  has been completed. 
In addition, one can expect a rich spectrum of states for medium and heavy mass nuclei.

\section{\label{sec:summary} Summary and conclusions}

We have calculated the spectra of $\eta_c$- and $J/\Psi$-nucleus bound states for various  nuclei.  
The meson-nucleus potentials were calculated using a local density approximation, by including only 
the $DD^{*}$ ($D\bar{D}$) meson loop in the $\eta_c$ ($J/\Psi$) self-energy.
The nuclear density distributions, as well as the in-medium $D$ and $D^{*}$ meson masses in nuclear matter were calculated within the QMC model. Then, using these potentials in nuclei, we solved the 
Klein-Gordon  equation and obtained meson--nucleus bound state energies. 
Our results show that one should expect the $\eta_c$ and $J/\Psi$ to form bound states for all the 
nuclei studied, even though  the precise values of the bound state energies are dependent on the 
cutoff mass values used in the form factors. The sensitivity of our results to the cutoff parameter 
used has also been explored. 

\section{Acknowledgments}

This work was partially supported by Conselho Nacional de Desenvolvimento 
Cient\'{i}fico e Tecnol\'{o}gico (CNPq), process Nos.~313063/2018-4 (KT), 
426150/2018-0 (KT) and 309262/2019-4 (GK), and Funda\c{c}\~{a}o 
de Amparo \`{a} Pesquisa do Estado de S\~{a}o Paulo (FAPESP) process 
Nos.~2019/00763-0 (KT), 64898/2014-5 (KT) and 2013/01907-0 (GK). The work 
is also part of the project Instituto Nacional de Ci\^{e}ncia e Tecnologia 
-- Nuclear Physics and Applications (INCT-FNA), process No.~464898/2014-5 
(KT, GK). It was also supported by the Australian Research Council through 
DP180100497 (AWT).

\end{document}